# Regarding the "Comments on 'Near-field interference for the unidirectional excitation of electromagnetic guided modes'" by Lee et al.

Francisco J. Rodríguez-Fortuño[1,2], Giuseppe Marino[1], Pavel Ginzburg[1], Daniel O'Connor[1], Alejandro Martínez[2], Gregory A. Wurtz[1] and Anatoly V. Zayats[1]

[1] Department of Physics, King's College London, Strand, London WC2R 2LS, UK
[2] Nanophotonics Technology Center, Universitat Politècnica de València, Valencia 46022, Spain

We are very sorry about the misunderstanding caused by comment [1] on our article [2]. We would like to emphasise that the vectorial near-field interference effect [2] is a fundamental physical process that is valid for all kinds of waves, both photonic and plasmonic and at all frequencies of the field (nonresonant) as discussed in detail in [2]. The discovery of this effect is the main topic of the paper. The effect has also recently been experimentally validated for surface plasmon polariton excitation by gratings (see Supplementary On-line Material (SOM) in [2]), photonic mode excitation by gratings [3], circularly-polarised dipole emulated by a nanoparticle [3], directional excitation of radio-frequency guided modes in hyperbolic metamaterials [4], as well as directional coupling in Si photonic waveguides [5].

The experimental illustration of the effect in [2] was performed using a slit as the equivalent of a 2D dipole. The analogy with the effect discussed in [1] is, therefore, limited to this practical particularity, the underlying concept discussed being clearly different. In particular, comment [1] points out the importance of the dipole moment induced along the slit (Y-direction of the co-ordinate system). We are aware that this is necessary for a complete analysis of this particular slit experiment, but it is not relevant to our general theory of vectorial near field interference. We have demonstrated unidirectional excitation of guided modes using dipoles circularly polarized in the XZ plane (see Figs. 1, 2 and S3 in [2]), thereby showing that the dipole $p_y$ is not a necessary ingredient. It is also clearly stated in [2] that grazing incidence is a fundamental requirement used by us in the experiment to achieve the effective dipole oscillating in the XZ plane (with no claim that there are no dipoles directed along Y) and not for practical difficulties as claimed in [1].

Further, we would like to point out that the symmetry considerations described in [1] to explain unidirectionality were already carefully considered in our original paper (see [2] and SOM in [2]).

Finally, we would like to thank the authors of [1] for bringing our attention to their article (ref. 3 in [1]) which we were unaware of at the time of publication of our paper [2].